\documentclass[twocolumn,journal]{IEEEtran}
\usepackage[LGR,T1]{fontenc}
\usepackage[latin9]{inputenc}
\usepackage{float}
\usepackage{textcomp}
\usepackage{amsmath}
\usepackage{graphicx}
\usepackage{setspace}
\usepackage[unicode=true,
 bookmarks=true,bookmarksnumbered=true,bookmarksopen=true,bookmarksopenlevel=1,
 breaklinks=false,pdfborder={0 0 0},pdfborderstyle={},backref=false,colorlinks=false]
 {hyperref}
\hypersetup{pdftitle={Your Title},
 pdfauthor={Your Name},
 pdfpagelayout=OneColumn, pdfnewwindow=true, pdfstartview=XYZ, plainpages=false}

\makeatletter

\DeclareRobustCommand{\greektext}{%
  \fontencoding{LGR}\selectfont\def\encodingdefault{LGR}}
\DeclareRobustCommand{\textgreek}[1]{\leavevmode{\greektext #1}}
\ProvideTextCommand{\~}{LGR}[1]{\char126#1}

\providecommand{\tabularnewline}{\\}

\let\oldforeign@language\foreign@language
\DeclareRobustCommand{\foreign@language}[1]{%
  \lowercase{\oldforeign@language{#1}}}

\usepackage[caption=false,font=footnotesize]{subfig}

\makeatother

\begin{document}
\title{Depth-Optimized Delay-Aware Tree (DO-DAT) for Virtual Network Function
Placement}
\author{Dimitrios Michael Manias, Hassan Hawilo, Manar Jammal and Abdallah
Shami\thanks{Dimitrios Michael Manias, Hassan Hawilo and Abdallah Shami are with
the Department of Electrical and Computer Engineering at The University
of Western Ontario, London, Canada, e-mail: \{dmanias3, hhawilo, Abdallah.shami\}@uwo.ca.
Manar Jammal is with the School of IT at York University, Toronto,
Canada, e-mail: mjammal@yorku.ca}}
\markboth{To Appear in: IEEE Networking Letters}{Your Name \MakeLowercase{\emph{et al.}}: Your Title}
\maketitle
\begin{abstract}
With the constant increase in demand for data connectivity, network
service providers are faced with the task of reducing their capital
and operational expenses while ensuring continual improvements to
network performance. Although Network Function Virtualization (NFV)
has been identified as a solution, several challenges must be addressed
to ensure its feasibility. In this paper, we present a machine learning-based
solution to the Virtual Network Function (VNF) placement problem.
This paper proposes the Depth-Optimized Delay-Aware Tree (DO-DAT)
model by using the particle swarm optimization technique to optimize
decision tree hyper-parameters. Using the Evolved Packet Core (EPC)
as a use case, we evaluate the performance of the model and compare
it to a previously proposed model and a heuristic placement strategy.
\end{abstract}


\section{Introduction}

\IEEEPARstart{W}{}ith network connectivity demands at an all-time
high and continuing to increase, Network Service Providers (NSPs)
are tasked with the challenge of accommodating additional bandwidth
requests on their networks while concurrently maintaining or improving
their Quality of Service (QoS). To adapt their networks to accommodate
this demand, NSPs must create a network with increased flexibility,
portability, and scalability. The concept of Network Function Virtualization
(NFV) has been proposed as a candidate solution for addressing these
challenges. NFV architecture isolates network functions and executes
them as software-based applications independently from the underlying
hardware \cite{r1}. By abstracting the individual network functions
from their underlying hardware and creating Virtual Network Functions
(VNFs), NSPs may experience a reduction in capital and operational
expenditures, and an increase in operational efficiencies \cite{r2}.

NFV technology, however, is not without its own challenges, including
performance, availability, and reliability. NSPs are obliged to adhere
to specific standards when delivering a service to a customer. These
standards are outlined through QoS guarantees, performance metrics,
and thresholds pertaining to jitter, packet loss, delay, and availability.
When evaluating the feasibility of an NFV-enabled network, adherence
to QoS guarantees is essential and must be considered.

One of the key metrics outlined in QoS guarantees is performance,
which can be described by different metrics such as delay or availability
and can pertain to an individual VNF instance or a set of interconnected
VNF instances known as a Service Function Chain (SFC).

Our previous work presents the Delay Aware Tree (DAT), which uses
a decision tree to address the NP-Hard VNF Placement Problem \cite{r3-1}.
The DAT shows promising results when compared to current heuristic
solutions. However, the DAT placement strategy, on average, produces
34 \textgreek{m}s of additional delay per computational path when
compared to current heuristics due to sub-optimal fitting. When considering
the incoming adoption of 5G networks and the new ultra-low latency
requirements (<1ms) in industrial internet of things use cases, this
added delay hinders the adoption of the DAT. As such, in this work,
the maximum depth hyperparameter (related to fitting) of the DAT is
optimized in an effort to improve the delay observed across all computational
paths and outperform current heuristics. To optimize the maximum depth
of the DAT, we propose the optimization of a performance-based objective
function, which considers both the delay and QoS guarantees when evaluating
the fitness of a set of hyperparameter values.

In order to illustrate the proposed solution, the virtual Evolved
Packet Core (vEPC) is selected as a use case; however, the solution
presented in this paper is generalizable to any SFC. There are four
VNF instances forming the SFC for vEPC being: the Home Subscriber
Service (HSS), the Mobility Management Entity (MME), the Serving Gateway
(SGW), and the Packet Data Network Gateway (PGW).

The remainder of this paper is structured as follows. Section II discusses
the state-of-the-art. Section III outlines the methodology. Section
IV presents and analyzes the results obtained. Finally, Section V
concludes the paper.

\section{Related Work}

There has been significant work in the field of VNF placement in recent
years. Some methods used to address the VNF placement problem include
optimization problem formulations \cite{mr21}, latency-aware placement
schemes \cite{mr22}, Monte-Carlo tree-based chaining algorithms \cite{mr23},
and matching theory approaches \cite{mr24}. The abovementioned works
however, are not capable of learning from historical observations;
to address this inadequacy, ML-based solutions are explored. Wahab
\textit{et al.} \cite{r4-1} propose an ML approach for efficient
placement and adjustment of VNFs and minimizing operational costs
while considering capacity and efficiency constraints. Khezri \textit{et
al.} \cite{r4} propose a deep Q-learning model considering the reliability
requirements of a given service function chain. Zhang \textit{et al.}
\cite{r5} propose an intelligent cloud resource manager that uses
deep reinforcement learning when mapping services and applications
to resource pools. Sun \textit{et al.} \cite{r6} propose Q-learning
as a method of addressing the time-accuracy tradeoff between heuristic
and optimization models. Khoshkholghi \textit{et al.} \cite{key-1}\textit{
}propose a genetic algorithm with the objective of minimizing a resource-based
cost function. Compared to these studies, our work advances the state-of-the-art
as we capture carrier-grade functionality constraints (\textit{i.e}.
availability) as well as the dependency constraints while simultaneously
generating placements that produce multiple computational paths (CPs)
(\textit{i.e.} multiple components serving the same SFC) which enables
the minimization of end-to-end SFC delay as well as enhanced availability.
Our work also considers HyperParameter Optimization (HPO) and analyzes
its effect on the overall performance of the model.

HPO is used to improve the performance of ML algorithms. The tuning
and optimization of tree-based machine learning models has been explored
using searches, heuristics and metaheuristics \cite{r7}, \cite{r8},
visual methods \cite{r9}, and Bayesian optimization \cite{r10}.
The main metric for assessing performance in these works has been
predictive accuracy.

This work extends our previous work by introducing a method of optimizing
the performance of the DAT through HPO. Due to the nature of this
multi-class, multi-output classification problem, predictive accuracy
is not sufficient as a metric for evaluating our model. The main contributions
of this paper include: a domain-based HPO model, which optimizes the
maximum depth parameter of the DAT using the meta-heuristic Particle
Swarm Optimization (PSO) technique, the introduction of a regularization
term, which severely penalizes invalid placement predictions, and
the creation of the Depth-Optimized Delay-Aware Tree (DO-DAT), which
exhibits improved performance compared to placement strategies published
in literature and facilitates automation in NFV management and orchestration.

\section{Methodology}

The following section outlines the various stages leading to the development
of the DO-DAT.

\subsection{Problem Formulation}

The problem formulation for this work is conducted in a two-fold manner,
the first dealing with the problem formulation of the DAT and the
second dealing with the problem formulation of the PSO depth optimization.

\subsubsection{DAT}

The methodology behind the construction of the DAT, as defined by
our previous work \cite{r3-1}, takes the previous placements made
by the near-optimal heuristic BACON algorithm \cite{r22}. Inherently,
the problem formulation for the DAT follows the MILP problem formulation
for the BACON algorithm outlined in the work of Hawilo \textit{et
al.} \cite{r22} and constructs a dataset that is used to train the
DAT. The BACON problem formulation has the objective of minimizing
the delay experienced by two dependent VNFs forming an SFC. To capture
the carrier-grade requirements associated with this technology, several
constraints were included in the problem formulation, including capacity
constraints (placement cannot exceed computational resource capacity),
network-delay constraint (placement cannot violate latency requirement),
availability constraint (placement cannot violate co-location and
anti-location requirements), redundancy constraint (placement must
improve availability through the placement of redundant components),
and dependency constraint (placement must ensure that dependent VNFs
forming an SFC are placed in a manner which enables the execution
of the SFC).

\subsubsection{PSO depth optimization}

The PSO depth optimization is conducted through the development of
a unique optimization function related to the domain of the NFV-enabled
network. By adopting this process, it is possible to move past the
point of matching the performance of the BACON algorithm and instead
focus on the continual development of the predictive placement model
as a whole. The PSO optimization takes place once during the initial
construction of the DO-DAT.

When considering the construction of a decision tree, the maximum
depth of the tree has been identified as a key hyperparameter in the
overall fitting of the model. In an effort to prevent over and underfitting,
this work presents a joint optimization objective that considers both
the average delay across all CPs of a predicted placement as well
as a penalty factor related to fitting. In the previous construction
of the DAT, improper model fitting manifested itself through invalid
predicted placements, which were instances where the constraints imposed
on the initial BACON problem formulation were not captured by the
DAT and therefore resulted in predicted placements which were considered
invalid. The penalty factor term operates like a regularization term
in the objective function penalizing invalid predicted placements
during the training phase of the model.

The formulation of the multi-objective optimization problem consolidated
into a single objective function is defined below.

The hyperparameter set is defined in (1)

\begin{equation}
\{h\}=\left\{ maxDepth\right\} 
\end{equation}

The objective of the optimization is to minimize the delay across
all CPs as well as the number of invalid predicted placements. Let
\textit{i} represent the trial number and \textit{j} represent the
CP; the average delay across all CPs can be defined as:

\begin{equation}
avg_{delay,CP}=\frac{\sum_{i=1}^{n}\left[\frac{\sum_{j=1}^{k}delay_{CP_{j}}}{k}\right]}{n}
\end{equation}

Where \textit{n} is the total number of trials and \textit{k} is the
total number of CPs.

The regularization term used is expressed through (3) where \textit{ip}
represents invalid placement predictions. In order to be effective,
the regularization term must have an equal order of magnitude compared
to the quantity being regularized; since the average delay per CP
value is in the order of thousands of microseconds, 1000 was selected
as a coefficient to ensure that the regularization term firstly is
of equal magnitude and additionally has a severe penalty on the overall
objective.

\begin{equation}
regTerm=1000*\log_{2}(ip+1)
\end{equation}

It is evident that as the number of invalid predictions approaches
zero, so does this regularization term suggesting that in the ideal
case where there are zero invalid predictions, the effect of this
regularization term is zero, as seen in (4).

\begin{equation}
\lim_{ip\rightarrow0}1000*\log_{2}(ip+1)=0
\end{equation}

By combining (2) and (3) into a single, equally-weighted objective
function, the following is obtained:

\begin{equation}
O_{PSO}=avg_{delay,CP}+regTerm
\end{equation}

The evaluation criterion considers the objective (5) evaluated on
a model with a given hyperparameter value $(model(h))$ across the
training and validation sets $T_{s}$and $V_{s}$.

\[
E(O_{PSO},model(h),T_{s},V_{s})
\]

Considering cross-validation, where\textit{ b} represents the number
of folds, the following is the function to be optimized.

\[
P(h)_{PSO}=\frac{1}{b}\sum_{i=1}^{b}E(O_{PSO},model(h),T_{s}^{(i)},V_{s}^{(i)})
\]

Finally, since this is a cost function, the optimization problem objective
is expressed in (6) as a minimization:

\begin{equation}
minimize\:P(h)_{PSO}
\end{equation}

When considering the above optimization problem, the possible range
of hyperparameter values is constrained to the functional range of
the system. In this problem, the functional range is defined as the
range where the number of invalid placement predictions falls below
a specified error threshold and when it achieves steady-state across
ten depth iterations. The constraint on the hyperparameter values
is defined by:

\[
a_{1}\leq h\leq a_{2}
\]
where the functional range is defined on the interval $[a_{1},a_{2}]$.

\subsection{Data Generation}

In order to generate the training and testing datasets, initial network
topologies are constructed. These topologies are structured as 3-tier
data centers to simulate the NFV-enabled network environments operated
by NSPs, as defined in our previous work \cite{r3-1}. In order to
generate the topology, an initial number of network servers and vEPC
VNF instances are selected. For each server-instance permutation,
10,000 topologies are generated with differing network conditions
and the respective VNF instances are placed using the BACON algorithm.
All the network parameters (\textit{i.e.} delays, resources, tolerances,
etc.) were simulated by sampling published distributions from NSP
datacenters such as Microsoft \cite{key-8} and Intel \cite{key-9}.
By sampling these distributions, we can ensure our work is generalizable
and applicable to real-world environments.

In this work, two different server-instance permutations are considered,
and 20,000 topologies are generated. The first contains 6 VNF instances
to be placed on 15 network servers. Taking into consideration the
distribution of VNF instances, there are a total of 4 different CPs
available for this topology. The second considers the placement of
10 VNF instances on 30 network servers and 36 CPs.

\subsection{Data Analysis}

\begin{singlespace}
Upon the creation of the various topologies through the data generation
and initial placements using the BACON algorithm, the next stage in
the methodology relates to the feature extraction. In order to predict
the placement of VNF instances on network servers, a snapshot of the
network conditions is taken and used as input features to the model.
The output labels are the placements of the components on the network
servers; as previously stated, this is a multi-class, multi-output
problem; therefore, there is a set of outputs predictions, each with
their respective set of possible labels. Given that, \textit{s} represents
a network server, and \textit{v} represents an instance to be placed,
the following holds true:
\end{singlespace}

\begin{align*}
outputs & =\left\{ v_{1},v_{2},...,v_{n}\right\} \\
labels & =\left\{ s_{1},s_{2},...,s_{n}\right\} 
\end{align*}

From the network snapshot, several features are extracted, including
instance resource requirements, server resource capacity, delay tolerance
between interdependent instances, delay between server, and instance
dependency levels.

\subsection{Model Construction}

The construction of the DO-DAT follows a 3 step process. The first
stage involves the determination of the range of under/overfitting
with respect to the tree depth. PSO optimization is initially run
given a range of {[}2,100{]} for the maximum depth hyperparameter.
The value of 100 is selected as the upper bound for the range of values
that the maximum depth hyperparameter can assume as a benchmark to
limit the initial search space; if an optimal solution is not achieved
in the defined search space, the upper bound is increased by a factor
of 2 until an optimal solution is found. Since the evidence of under/overfitting
in the DAT was manifested through the number of invalid placement
predictions, the goal of this PSO optimization stage is to determine
the depth which minimizes their occurrence.

The result from the first stage is then used to determine the range
of values to be further considered. The range of values is determined
by considering when the number of invalid placement predictions falls
below an initial error threshold set at 7.5\% (10\% is the pre-defined
maximum tolerable placement error, by setting the initial threshold
to 7.5\%, we can be more conservative with our placements and reduce
the search space for the subsequent steps) and when steady-state is
reached, meaning that there is no further improvement observed for
several iterations. The optimization performed in the second stage
considers the entire objective function, (6) evaluated across the
previously determined range. The result of this optimization shows
the effect of the range of depths on the joint consideration of invalid
predictions and delay.

The final stage of the construction of the DO-DAT is to identify the
optimal tree depth obtained from the previous stage and construct
the model using this hyperparameter value.

\section{Results and Analysis}

The following is a presentation of the results obtained as well as
an analysis of their implications on the DO-DAT. The generation of
the dataset, data processing and ML models are implemented using Python
on a PC with an Intel Core\texttrademark{} i7-8700 CPU @ 3.20 GHz
CPU, 32 GB RAM, and an NVIDIA GeForce GTX 1050 Ti GPU.

\subsection{Functional Range}

The first set of results pertains to the PSO optimization and its
effectiveness in presenting the optimal value of the tree depth. Fig.
1 displays the effect of varying the depth of the tree on the number
of invalid placement predictions.

\begin{figure}[h]
\centerline{\includegraphics[width=7cm,height=3.5cm]{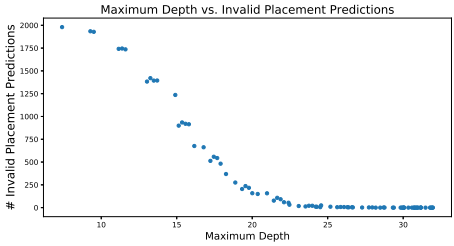}}\caption{Effect of max depth on invalid predictions}

\end{figure}
As seen, the number of invalid placement predictions decreases while
the tree depth is less than 25 and stabilizes at this minimum value
of 0 while the tree depth is greater than 25. At a max depth of 20,
the number of invalid placement predictions (error rate) is 7.5\%.
Since steady-state is observed beyond 25, 10 additional tree depths
were considered to evaluate the effect of the overfitting on the optimization.
Therefore, the range of tree depths spanning {[}20,35{]} is selected
as the functional range of the first stage and is further evaluated
by taking into consideration the full objective function (6).

\subsection{Optimal Depth}

Results from the optimization of the functional range of interest
are presented in Fig. 2. From this figure, we can see the objective
function \textit{P(h)} is decreasing on the interval {[}20,28{]} and
plateaus on the interval {[}29,35{]}. The interval {[}28,30{]} represents
the interface between under and overfitting of the DAT, and therefore,
since there is no further significant improvement on the interval
{[}29,35{]}, the optimal depth is 29.

\begin{figure}[h]
\centerline{\includegraphics[width=7cm,height=3.5cm]{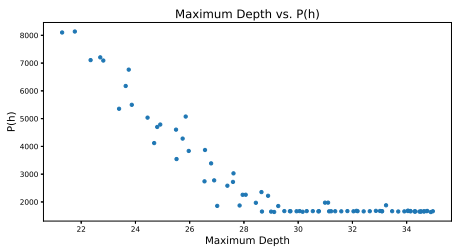}}\caption{Effect of depth on \textit{P(h)}}

\end{figure}

\subsection{Performance Comparison}

The following figures compare the placement of BACON, DAT, and DO-DAT.
Fig. 3 illustrates the delay across the various CPs in the small scale
network. As observed in this Fig. 3, DO-DAT exhibits improved performance
when compared to both the BACON and DAT placement strategies.

\begin{figure}[h]
\centerline{\includegraphics[width=7cm,height=3.5cm]{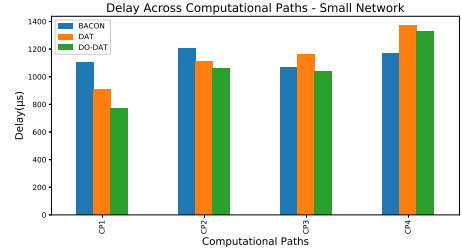}}\caption{Delay across computational paths of the small network}
\end{figure}
Fig. 4 displays the delay experienced between interconnected dependent
instances, forming a vEPC SFC.

\begin{figure}[H]
\centerline{\includegraphics[width=7cm,height=3.5cm]{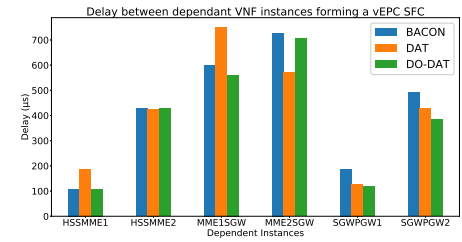}}\caption{Delay between dependent VNF instances forming a vEPC SFC}

\end{figure}
As seen in Fig. 4, the DO-DAT is the best performing placement strategy
as it has successfully placed the VNF instances with less delay exhibited
between dependent VNF instances. These results can be further extended
to the second network topology, as expressed in Fig. 5, where it can
be seen that DO-DAT, when considered across all CPs, produces more
paths with less delay compared to the other placement strategies.
This is further illustrated in Table 1, where the ratios of which
strategy produces a placement with the least delay per each of the
36 CPs seen in Fig. 5 are listed. In all cases, the DO-DAT outperforms
the other two strategies and establishes itself as the clear winner
when considering the reduction of end-to-end delay across all CPs.

\begin{figure}[h]
\begin{centering}
\centerline{\includegraphics[width=8.5cm]{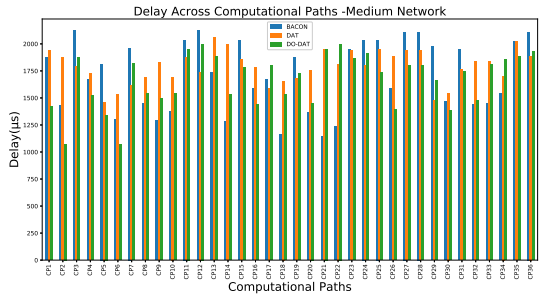}}\caption{Delay across computational paths medium network}
\par\end{centering}
\end{figure}
\begin{table}[H]

\caption{Placement strategy producing the lowest delay per CP}

\begin{centering}
\begin{tabular}{|c|c|}
\hline 
{\scriptsize{}Ratio Description} & {\scriptsize{}Ratio}\tabularnewline
\hline 
\hline 
{\scriptsize{}BACON vs. DAT vs. DO-DAT} & {\scriptsize{}13:9:14}\tabularnewline
\hline 
{\scriptsize{}BACON vs. DO-DAT} & {\scriptsize{}13:23}\tabularnewline
\hline 
{\scriptsize{}DAT vs. DO-DAT} & {\scriptsize{}12:24}\tabularnewline
\hline 
\end{tabular}
\par\end{centering}
\end{table}
This is reinforced when considering Fig. 6, which shows a Probability
Density Function (PDF) of the difference between the DO-DAT and BACON
algorithms in terms of placement delay; a similar comparison between
BACON and DAT was conducted in our previous work \cite{r3-1}. By
calculating the difference in delay across every CP placement, we
can determine the probability of the DO-DAT outperforming BACON. The
mean in Fig. 6 is -10\textgreek{m}s; therefore, on average, DO-DAT
provides a CP with 10\textgreek{m}s less of a delay when compared
to BACON. This is an improvement on our previous work related to DAT,
which on average had 34\textgreek{m}s more delay. The work presented
in the paper effectively improves the placement of VNF instances by
44\textgreek{m}s on average.

\begin{figure}[h]
\begin{centering}
\centerline{\includegraphics[width=6.5cm]{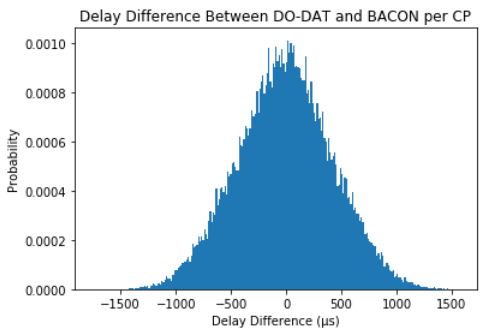}}\caption{Delay difference between DO-DAT and BACON}
\par\end{centering}
\end{figure}

\subsection{Runtime Complexity}

One of the benefits of the use of ML in networks is the reduction
of system complexity. This is evident through the runtime complexity
analysis of our proposed model. When considering the BACON algorithm,
it has a runtime complexity of $O(\frac{s^{3}-s^{2}}{2})$ where \textit{s}
denotes the number of available servers in the network \cite{r22}.
Our previous work outlines the time complexity of constructing a decision
tree as $O(n_{features}*n_{samples}*\log n_{samples})$ when creating
the tree and $O(\log n_{samples})$ when executing a query \cite{r25}.
Additionally, the DO-DAT has an additional offline optimization component
with the complexity defined by $O(n^{2}t)$ where \textit{n} denotes
the population and \textit{t} the iteration \cite{r26}. Since the
building of the tree and the optimization are completed entirely offline,
only the querying phase is considered during the runtime analysis. 

\section{Conclusions and Future Work}

The work presented in this paper described a key step towards an implementable,
intelligent, and delay-aware VNF placement strategy. Through the optimization
of the max tree depth, we addressed the under/overfitting phenomenon,
which plagues large decision trees and negatively impacts performance.
The DO-DAT uses ML and PSO to provide an effective, real-time placement
solution, which outperforms existing placement strategies and improves
QoS through the reduction of the delay between VNF instances, forming
an SFC. Future work will consider the use of ML to address additional
services offered by the VNF orchestrator.

\end{document}